\documentclass[twocolumn,showkeys,showpacs,preprintnumbers,prd,superscriptaddress,nofootinbib]{revtex4-1}
\usepackage{verbatim}
\usepackage[T1]{fontenc}
\usepackage[utf8]{inputenc}
\usepackage[american]{babel}
\usepackage{epsfig}
\usepackage{booktabs}
\usepackage{multirow}
\usepackage{dcolumn}
\usepackage{amsmath}
\usepackage{mathtools}
\usepackage{amsfonts}
\usepackage{amssymb}
\usepackage{ulem}
\usepackage{epstopdf}
\usepackage{bm}
\usepackage{siunitx}
\usepackage{braket}
\usepackage{enumitem}
\usepackage{soul}
\usepackage[table]{xcolor}
\usepackage{color}
\usepackage{transparent}
\usepackage{pifont}
\usepackage{enumitem}

\definecolor{navyblue}{rgb}{0.0, 0.0, 0.5}
\definecolor{royalblue}{rgb}{0.25, 0.41, 0.88}
\definecolor{cadmiumgreen}{rgb}{0.0, 0.42, 0.24}
\definecolor{blue-violet}{rgb}{0.54, 0.17, 0.89}
\definecolor{darkviolet}{rgb}{0.58, 0.0, 0.83}
\definecolor{orange(colorwheel)}{rgb}{1.0, 0.5, 0.0}

\usepackage{hyperref}
\hypersetup{
    colorlinks=true, 
    linkcolor=royalblue, 
    citecolor=magenta}

\usepackage{booktabs}
\usepackage{multirow}
\usepackage{dcolumn}
\usepackage{colortbl}

\makeatletter\let\expandableinput\@@input\makeatother

\begin{document}

\title{Quantifying the $S_8$ tension and evidence for interacting dark energy from redshift-space distortion measurements}

\author{Miguel A. Sabogal}
\email{miguel.sabogal@ufrgs.br}
\affiliation{Instituto de F\'{i}sica, Universidade Federal do Rio Grande do Sul, 91501-970 Porto Alegre RS, Brazil}

\author{Emanuelly Silva}
\email{emanuelly.santos@ufrgs.br}
\affiliation{Instituto de F\'{i}sica, Universidade Federal do Rio Grande do Sul, 91501-970 Porto Alegre RS, Brazil}

\author{Rafael C. Nunes}
\email{rafadcnunes@gmail.com}
\affiliation{Instituto de F\'{i}sica, Universidade Federal do Rio Grande do Sul, 91501-970 Porto Alegre RS, Brazil}
\affiliation{Divisão de Astrofísica, Instituto Nacional de Pesquisas Espaciais, Avenida dos Astronautas 1758, São José dos Campos, 12227-010, São Paulo, Brazil}

\author{Suresh Kumar}
\email{suresh.kumar@plaksha.edu.in}
\affiliation{Data Science Institute, Plaksha University, Mohali, Punjab-140306, India}

\author{Eleonora Di Valentino}
\email{e.divalentino@sheffield.ac.uk}
\affiliation{School of Mathematics and Statistics, University of Sheffield, Hounsfield Road, Sheffield S3 7RH, United Kingdom}

\author{William Giar\`e}
\email{w.giare@sheffield.ac.uk}
\affiliation{School of Mathematics and Statistics, University of Sheffield, Hounsfield Road, Sheffield S3 7RH, United Kingdom}

\begin{abstract}
\noindent
In recent years, Cosmic Microwave Background (CMB) observations, Weak Lensing surveys, and $f\sigma_8(z)$ measurements from Redshift-Space Distortions (RSD) have revealed a significant ($\sim$3$-$5$\sigma$) discrepancy in the inferred value of the matter clustering parameter $S_8$. In this work, we investigate the implications of RSD for a cosmological framework 
postulating an interaction between Dark Energy (DE) and Dark Matter (DM). We explore scenarios where DM can transfer energy-momentum to DE or vice versa. The energy-momentum flow is characterized by the strength and the sign of the coupling parameter $\xi$. Our baseline analysis combines RSD measurements with the latest data from Baryon Acoustic Oscillations (BAO) observed by DESI, Type Ia Supernovae from the PantheonPlus sample, and CMB data from Planck. We demonstrate that RSD measurements provide significant additional information imposing new and strong upper bounds on possible interaction in the dark sector. Models with $\xi > 0$ can effectively alleviate the tension in $S_8$, presenting them as compelling alternatives.
\end{abstract}



\maketitle

\section{Introduction}
\label{sec:introduction}
Nowadays it is widely accepted that an exotic source of energy with negative pressure known as Dark Energy (DE) is responsible for the accelerated expansion of the universe~\cite{riess1998observational,perlmutter1999measurements} and somehow could be linked with its structure formation history.\footnote{Alternatively, modifications to general relativity can also be considered to explain the late-time acceleration of the universe. See~\cite{Ishak_2018,saridakis2023,Heisenberg_2019} for a review.} In the standard model of cosmology ($\Lambda$CDM), DE is modeled as a cosmological constant ($\Lambda$), independent of the Cold Dark Matter (CDM) content. While the $\Lambda$CDM model can almost perfectly reproduce a large number of observations~\cite{Planck:2018_1,Planck:2018_2,Planck:2018_3,Mossa_2020,Atacama_2020,eBOSS_2020,SPT-3G:2022hvq}, the increasing precision of observational data has revealed some discrepancies or ``tensions" at the core of the $\Lambda$CDM framework. Specifically, there is a tension at more than $5\sigma$ between the early and late-time estimates of the present-day expansion rate of the universe, known as the Hubble constant ($H_{0}$) tension~\cite{Riess_H0_2021,Verde:2019ivm,Knox:2019rjx,DiValentino:2020zio,DiValentino:2021izs,DiValentino:2022fjm,Kamionkowski:2022pkx,Verde:2023lmm,DiValentino:2024yew,Breuval:2024lsv,Li:2024yoe,Murakami:2023xuy}.  
Likewise, cosmic shear surveys and Planck-2018 Cosmic Microwave Background (CMB) anisotropy measurements point towards yet another tension surrounding the value of the weighted amplitude of matter fluctuations $S_{8} = \sigma_{8} \sqrt{\Omega_{\rm m}/0.3}$ (where $\sigma_{8}$ is the amplitude of matter fluctuations on scales of $8h^{-1}\mathrm{Mpc}$, and $\Omega_{\rm m}$ is the present-day density parameter of matter) as inferred within $\Lambda$CDM, see e.g. Refs.~\cite{DES_2021,DiValentino:2020vvd,DiValentino:2018gcu,Kilo-DegreeSurvey:2023gfr,Troster:2019ean,Heymans:2020gsg,Dalal:2023olq,Chen:2024vvk,ACT:2024okh,DES:2024oud,Harnois-Deraps:2024ucb,Dvornik:2022xap}. This is known as the $S_8$ tension.
These two tensions have drawn the community's attention to investigate the extensions of the $\Lambda$CDM model which can explain these discrepancies (see the reviews~\cite{DiValentino:2021izs,Abdalla_2022,Perivolaropoulos_2022,Khalife:2023qbu,DiValentino:2024yew} and references therein).

The tension in $S_8$ is 
directly related to the formation and evolution of cosmic structures. The CMB anisotropy measurements from Planck-2018 in the $\Lambda$CDM model provide the best fit: $S_{8} = 0.834 \pm 0.016$~\cite{Planck:2018_2}. This estimate finds an increasing statistical tension of $2$-$3\sigma$ levels with cosmic shear measurements~\cite{DiValentino:2020vvd}. For Weak Gravitational Lensing (WGL), the Kilo-Degree Survey (KiDS-1000) analysis provided the constraint: $S_{8}=0.759^{+0.024}_{-0.021}$~\cite{KiDS:2020suj}, which is in a $3\sigma$ tension with Planck-2018; the Hyper Suprime-Cam Year 3 reported a $2\sigma$ tension ($S_{8} = 0.776^{+0.032}_{-0.033}$)~\cite{Dalal:2023olq}, and the most recent Dark Energy Survey Year 3 (DES-Y3) analyses measured $S_{8} = 0.759^{+0.025}_{-0.023}$~\cite{DES:2021bvc} with a $2.3\sigma$. It is worth mentioning that the analysis of galaxy clustering and WGL of the DES-Y3 data combining three two-point functions (3 $\times$ 2pt analysis) and galaxy positions (with an improvement in signal-to-noise relative to DES-Y1 by a factor of 2.1), gives $S_{8} = 0.776^{+0.017}_{-0.017}$~\cite{DES_2021} in a flat $\Lambda$CDM model. On the other hand, Redshift Space Distortion (RSD) data alone show a $3.1\sigma$ tension with the Planck-CMB results~\cite{Nunes_Vagnozzi_2021, Kazantzidis:2018rnb}. When combining weak gravitational lensing, real-space clustering, and RSD data, the authors of~\cite{Skara_2019} observed a significant increase in the tension related to the growth of structure, rising from $3.5\sigma$ (based on RSD data alone) to $6\sigma$ when including $E_{g}$ data as well. See \cite{Perivolaropoulos:2021jda} for a more discussion on this topic.


Unlike the discrepancy in $H_{0}$, the $S_{8}$ tension is not yet absolute; however, it reflects the trend of Large Scale Structure (LSS) data to lower $S_{8}$ values than those obtained with early time probes, which opens room for new physics beyond $\Lambda$CDM (if one excludes the systematic error hypothesis)~\cite{Abdalla_2022}. 

A somewhat crucial difference between the  $H_0$-tension and the $S_8$-tension is that, while for $H_0$ we can compare direct measurements in the late universe (which are notably independent of the cosmological model) with model-dependent estimates inferred from early universe observations, all measurements of $S_8$ are intrinsically model-dependent.
Therefore, even assuming the same $\Lambda$CDM model, early universe measurements with the CMB and late universe measurements with LSS are in disagreement. 
Numerous approaches have been proposed to address the $S_{8}$ tension, such as decaying dark matter~\cite{DiValentino_2017,FrancoAbellan_2020,He:2023dbn,Tanimura:2023bkh,Fuss:2024dam}, active and sterile neutrinos~\cite{Battye_2013,Feng_2017,Caramete:2013bua,Feng:2019jqa,Camarena:2024zck}, modified gravity models~\cite{Nesseris_2017,Marra_2021,Acero:2024jqe,deAraujo:2021cnd,Archidiacono:2022iuu,Uzan:2023dsk,Adi:2020qqf,DeFelice:2020prd,daSilva:2024oov,Luongo:2023arr,Nguyen:2023fip}, among many others~\cite{Meerburg:2014bpa,Hlozek:2014lca,Camera:2017tws,DiValentino:2019dzu,Moreno-Pulido:2020anb,Lucca:2021dxo,Cheek:2022yof,Basilakos:2023kvk,Toda:2024ncp,Akarsu:2024qsi,Terasawa:2024agq,Akarsu:2024eoo,Carrion:2024itc,Cheek:2024fyc,Chen:2024vuf,Chakraborty:2024pxy,Stahl:2024stz,Amon:2022azi}. 
A model that has been extensively studied in the literature and shown remarkable promise in resolving various issues currently associated with the standard model is the so-called Interacting Dark Energy (IDE)~\cite{Kumar:2016zpg,Murgia:2016ccp,Kumar:2017dnp,DiValentino:2017iww,Yang:2020uga,Kumar:2021eev,Forconi:2023hsj,Benisty:2024lmj,Pourtsidou:2016ico,DiValentino:2020vnx,DiValentino:2020leo,Nunes:2021zzi,Yang:2018uae,vonMarttens:2019ixw,Lucca:2020zjb,Zhai:2023yny,Bernui:2023byc,Hoerning:2023hks,Giare:2024ytc,Escamilla:2023shf,vanderWesthuizen:2023hcl,Silva:2024ift,DiValentino:2019ffd,Li:2024qso,Pooya:2024wsq,Halder:2024uao,Castello:2023zjr,Pan:2023mie,Yao:2023jau,Mishra:2023ueo,Nunes:2016dlj}, where a non-gravitational interaction between DE and DM is 
postulated
(see e.g. Ref.~\cite{Wang_2024} for a recent review). Recently, in Ref.~\cite{Giare:2024smz} it was shown by some of us that the new Baryon Acoustic Oscillations (BAO) measurements from the Dark Energy Spectroscopic Instrument (DESI), in conjunction with Planck-2018 CMB data, exhibit a preference for DE-DM interaction over $\Lambda$CDM while reducing the $H_{0}$-tension. In this paper, we explore in detail the role of RSD samples in this cosmological context, considering both scenarios where DM can transfer energy to DE and the opposite case. Our analysis offers fresh insights, setting new upper bounds on the interaction parameter in both cases—$\xi < 0$ (DM to DE) and $\xi > 0$ (DE to DM)—when RSD data are incorporated 
Additionally, we demonstrate that models with $\xi > 0$ could potentially alleviate the tension observed in $S_8$.

This paper is structured as follows: Sec.~\ref{model} introduces the IDE model. Sec.~\ref{data} describes the methodology and datasets used in this study. The results of the analyses are presented in Sec.~\ref{results}. Finally, Sec.~\ref{conclusions} presents the conclusions drawn from our analyses.

\section{Interacting Dark Energy}
\label{model}
To review the basic features of IDE models, we adopt a spatially flat Friedmann-Lemaître-Robertson-Walker (FLRW) metric. In the absence of interactions between DE and DM, their stress-energy tensors, denoted by $T^{\mu\nu}_{\rm x}$ for DE and $T^{\mu\nu}_{\rm c}$ for DM, respectively, are individually covariantly conserved. In IDE models, a parameterization is introduced in the conservation equations such that the individual stress-energy tensors are no longer conserved, but their sum remains conserved~\cite{Wang_2024}. Therefore, the evolution of the covariant derivatives of the stress-energy tensors for DE and DM can be expressed as:

\begin{eqnarray}
\label{eq:covariant}
\sum_{j} \nabla_\mu T_{j}^{\mu\nu} = 0, \hspace{0.5cm}   \nabla_\mu T_{j}^{\mu\nu}= \frac{Q_{j} u_{\rm c}^\nu}{a},
\end{eqnarray}\\
\noindent where the index $j$ runs over DE and DM, $a$ is the scale factor, $u_{\rm c}^\nu$ denotes the DM four-velocity vector, and $Q_{\rm c} = -Q_{\rm x} = Q$ is the DE-DM interaction rate with units of energy per volume per time.

Due to the unknown nature of the dark sector, one must phenomenologically select the functional form of $Q$. In this study, we adopt a well-established parametric choice from the literature, specifically $Q = \xi {\cal H} \rho_{\rm x}$~\cite{Gavela:2010tm,DiValentino:2019ffd,DiValentino:2019jae,Zhai_2023}, where $\xi$ is a dimensionless parameter that governs the strength of the interaction between DE and DM. Here, ${\cal H}$ denotes the conformal Hubble rate, and $\rho_{\rm x}$ represents the energy density of DE. In this parametrization, when $\xi < 0$, energy-momentum transfers from DM to DE, while it is the opposite for $\xi > 0$.

Within the context of linear perturbations, the presence of non-gravitational interaction between DE and DM also influences their evolution. Adopting the synchronous gauge, the perturbed FLRW metric at linear order is given by:
\begin{equation}
ds^2 = a^2 [-d\tau^2 + (\delta_{ij} + h_{ij}) dx^i dx^j],
\label{eq}
\end{equation}
where $\tau$ is the conformal time, and $h_{ij}$ represents the scalar metric perturbation.

The coupled system of linear Einstein-Boltzmann equations governing the evolution of DE and DM density perturbations, $\delta_j$, and velocity divergences, $\theta_j$, can be expressed as~\cite{V_liviita_2008,Gavela:2010tm,Nunes:2022bhn}:
\begin{subequations}
\begin{eqnarray}
\dot{\delta}_{\rm c} & = & -\theta_{\rm c} - \frac{1}{2} \dot{h} +  \frac{\xi\rho_{\rm x}}{\rho_{\rm c}} \left[ \mathcal{H}\left(\delta_{\rm x} - \delta_{\rm c}\right) + \frac{k v_T}{3} + \frac{\dot{h}}{6} \right] \, , \\
\dot{\theta}_{\rm c} & = & -\mathcal{H} \theta_{\rm c} \, , \\ 
\dot{\delta}_{\rm x} & = & -\left(1 + w_{\rm x}\right)\left[ \theta_{\rm x} + \left(1 - w_{\rm x}\right) \frac{9 \mathcal{H}^{2}}{k^2} \theta_{\rm x} + \frac{\dot{h}}{2}\right] \\
& & - 3\mathcal{H}\left(1 - w_{\rm x}\right)\delta_{\rm x} - \xi \left[ \frac{3 \mathcal{H}^{2}}{k^2}(1 - w_{\rm x}) + \frac{k v_T}{3} + \frac{\dot{h}}{6}\right] \, , \nonumber \\ 
\dot{\theta}_{\rm x} & = & 2 \mathcal{H} \theta_{\rm x} + \frac{k^2}{1 + w_{\rm x}} \delta_{\rm x} + \frac{\xi \mathcal{H}}{1 + w_{\rm x}} \left( 2\theta_{\rm x} - \theta_{\rm c} \right) \, .
\label{EB_4}
\end{eqnarray}
\label{E_boltzmann}
\end{subequations}

In the above equations, $h$ denotes the trace of the metric perturbation $h_{ij}$, dots represent derivatives with respect to conformal time, and $v_T$ represents the center of mass velocity for the total fluid, as required by gauge invariance arguments~\cite{Gavela:2010tm}. We assume that DE does not cluster, implying a DE sound speed squared: $c_{s}^2 = 1$. For a discussion on how the DE-DM interaction affects the initial conditions of the coupled system of Eqs.~\eqref{E_boltzmann}, refer to Refs.~\cite{Gavela:2010tm,Nunes:2022bhn}. Equation \eqref{EB_4} necessitates $w_{\rm x} \neq -1$ to avoid instabilities. It has been demonstrated that to prevent gravitational and non-adiabatic instabilities at early times, $(1 + w_{\rm x})$ and $\xi$ must have opposite signs~\cite{Gavela:2009cy,He:2008si}. Therefore, in our study, we choose $w_{\rm x} = -0.999$ or $w_{\rm x} = -1.001$, restricting $\xi \leq 0$ or $\xi \geq 0$, respectively.

Since the main equations have been outlined, now we can turn our attention to how the structure formation is affected, focusing in particular on linear scales. To efficiently measure the impact of the DE-DM interaction on the evolution of matter perturbations, its theoretical predictions must be compared with cosmological observables, e.g., RSD, which arise from velocity-induced distortions that occur when mapping from real-space to redshift-space, caused by the peculiar motions of objects along the line of sight. These distortions introduce anisotropies in their clustering patterns~\cite{Kaiser:1987qv}, and are dependent on the growth of structure, making RSD measurements sensitive to the combination $f \sigma_{8}(z)$ or equivalently $f(a) \sigma_8(a)$, where $\sigma_8(a)$ is the variance of the mass distribution smoothed on a sphere of radius $R = 8h^{-1}\,\mathrm{Mpc}$, while $f(a)$ is the logarithmic derivative of $D(a)=\delta_{\rm m}(a)/\delta_{\rm m}(1)$ with respect to the scale factor:
\begin{eqnarray}
f(a) \equiv \frac{d\ln D(a)}{d\ln a} \,,
\label{f(a)}
\end{eqnarray}
in which it holds $\rho_{\rm m} \delta_{\rm m} = \rho_{\rm b} \delta_{\rm b} + \rho_{\rm c} \delta_{\rm c}$ since DE only interacts with DM, and the baryonic ($\rho_{\rm b} \delta_{\rm b}$) sector remains untouched relative to $\Lambda$CDM.

\begin{figure}[tpb!]
   \centering
   \includegraphics[width=\columnwidth]{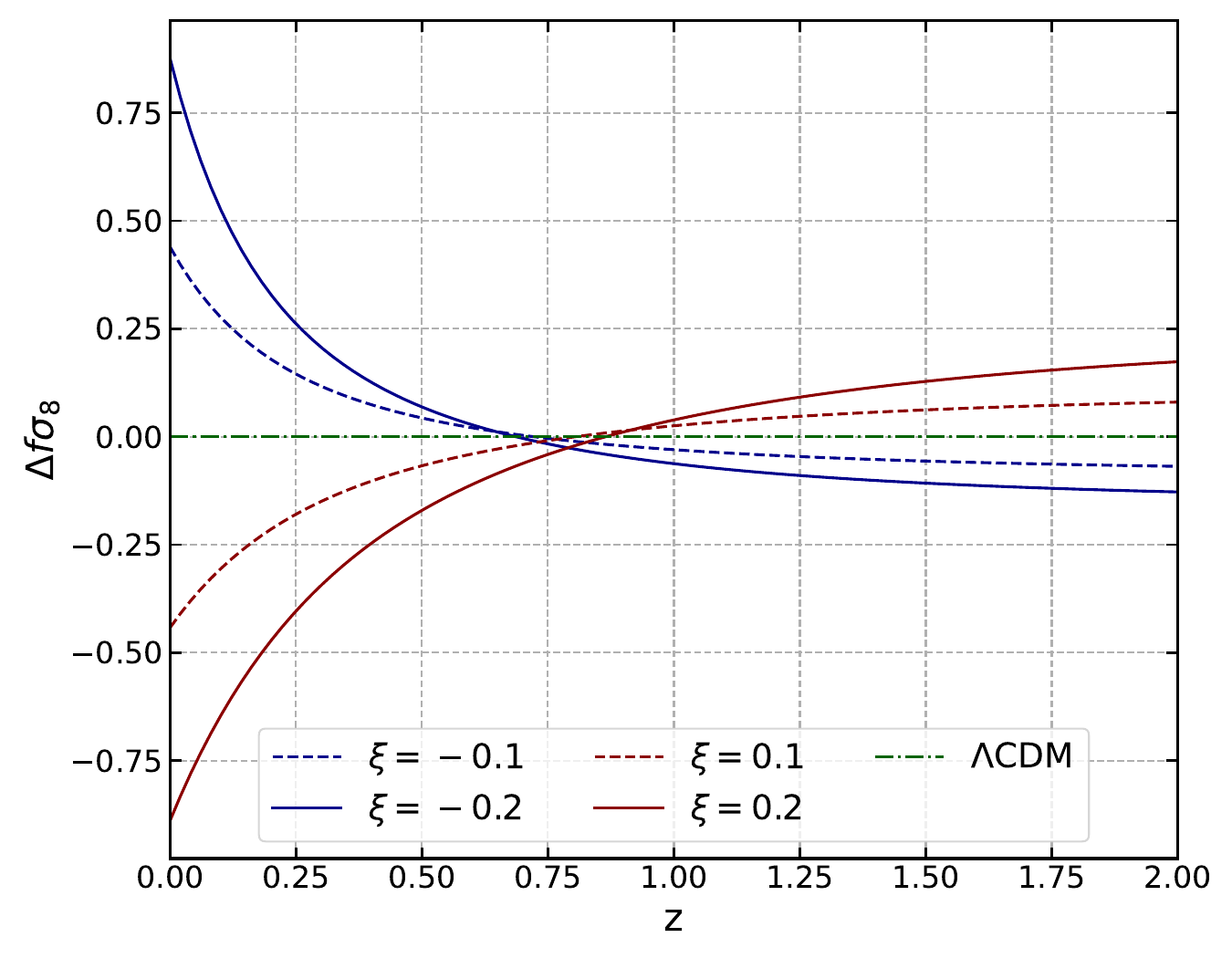} 
   \caption{
   Difference between the theoretical predictions of $f(z) \sigma_8(z)$ for $\Lambda$CDM and IDE models, normalized to their value in $\Lambda$CDM — see Eq.~\eqref{eq:Deltafsgima8}. The figure shows both negative and positive values of the coupling parameter $\xi$, as indicated in the legend.}
   
   \label{fig:fs8}
\end{figure}

In the present work, we assume that DE perturbations are only effective at large scales. Therefore, at sub-horizon scales where relativistic effects are negligible, gravity follows the usual Poisson equation, $-k^2 \Psi = 4 \pi G a^2 \rho_{\rm m} \delta_{\rm m}$. Following the procedure in Ref.~\cite{Silva:2024ift}, the evolution equation for $D(a)$ within the IDE scenario is given by:
\begin{equation}
\label{eq:D}
\begin{aligned}
\ddot{D} & + \mathcal{H} \left[ 1 + r_{\rm x m} \xi \right] \dot{D} \\
& - D \left[ \frac{3}{2} a^2 \rho_{\rm m} - r_{\rm x m} \xi \dot{\mathcal{H}} \right. \\
& \left. + r_{\rm x m} \xi \mathcal{H}^2 \left( \xi + 3 \omega_x + r_{\rm x m} \xi - 1 \right) \right] = 0,
\end{aligned}
\end{equation}
in which we denote $r_{\rm x m} = \rho_{\rm x}/\rho_{\rm m}$, and it should be noted that $\Lambda$CDM evolution is recovered for a null value of $\xi$.

The $\sigma_8(z)$ value is derived using the \texttt{CLASS} code~\cite{Blas:2011rf}, which directly captures model predictions in the transfer function for the modified Boltzmann equations of the model from Eqs. \eqref{E_boltzmann}. Consequently, when considering the total observable $f(z) \sigma_8(z)$, theoretical adjustments are anticipated for both quantities.

To quantitatively show the effects of the model on $f \sigma_8(z)$, we define

\begin{equation}
    \Delta f\sigma_8(z) = \frac{f\sigma_8(z)_{\mathrm{\Lambda CDM}} - f\sigma_8(z)_{\mathrm{IDE}}}{f\sigma_8(z)_{\mathrm{\Lambda CDM}}}.
    \label{eq:Deltafsgima8}
\end{equation}

Figure \ref{fig:fs8} illustrates how the quantity $f \sigma_8(z)$ is influenced by the strength and direction of the interaction between DE and DM. The red (blue) curves show a trend where $f \sigma_{8}$ is decreased (increased) relative to $\Lambda$CDM for $z < 0.75$ when energy flows from DE to DM (DM to DE), respectively. Conversely, for $z > 0.75$, the model predicts higher (lower) values compared to the $\Lambda$CDM model. Importantly, due to the tendency to amplify $f \sigma_{8}$ values in the nearby universe when energy transfer from DM to DE is enhanced, we anticipate that RSD data at $z < 0.1$ will impose strong bounds on scenarios with $\xi < 0$. During the construction of this plot, we kept all other common baseline parameters constant as per the CMB predictions, and varied only the coupling parameter $\xi$.

\section{Datasets and methodology}
\label{data}

We implemented the theoretical model in a modified version of the Boltzmann solver code \texttt{CLASS}~\cite{Blas:2011rf} and used the publicly available sampler \texttt{MontePython}~\cite{Brinckmann:2018cvx,Audren:2012wb} to perform Markov Chain Monte Carlo (MCMC) analyses, ensuring a Gelman-Rubin convergence criterion~\cite{Gelman_1992} of $R-1 \leq 10^{-2}$ in all runs. We assumed flat priors on the set of sampled cosmological parameters \{$\Omega_{\rm b} h^2$, $\Omega_{\rm c} h^2$, $\tau_{\rm reio}$, $100\theta_{\mathrm{s}}$, $\log(10^{10} A_{\mathrm{s}})$, $n_{\mathrm{s}}$, $\xi$\}, where the first six are baseline parameters within the $\Lambda$CDM context. Specifically, these parameters include the present-day physical density parameters of baryons ($\omega_b = \Omega_{\rm b} h^2$) and dark matter ($\omega_c = \Omega_{\rm c} h^2$), the optical depth of reionization ($\tau_{\rm reio}$), the angular size of the sound horizon at recombination ($\theta_{\mathrm{s}}$), the amplitude of the primordial scalar perturbation ($A_{\mathrm{s}}$), and the scalar spectral index ($n_{\mathrm{s}}$). The ranges of the priors are: $\omega_b \in [0.0, 1.0]$, $\omega_{\text{cdm}} \in [0.0, 1.0]$,
$100 \theta_s \in [1.03, 1.05]$, $\ln(10^{10} A_s) \in [3.0, 3.18]$, $n_s \in [0.9, 1.1]$, $\tau_{\text{reio}} \in [0.004, 0.125]$, and $\xi \in [0.0, 1.0]$ for the positive case or $\xi \in [-1.0, 0.0]$ for the negative case. In all analyses, we used the Python package \texttt{GetDist}\footnote{\url{https://github.com/cmbant/getdist}} 
to analyze the MCMC chains and extract the numerical results, as well as the 1D posteriors and 2D marginalized probability contours. \\

We summarize below the datasets used in our analysis: 
\begin{itemize}

\item \textit{Redshift Space Distortions} (\textbf{RSD}): 
We compile 20 measurements of $f\sigma_8(z)$ spanning the redshift range $0.02 < z < 1.944$, sourced from various surveys and references in the literature, as summarized in~\cite{Avila:2022xad}. The data points are selected as per the following criteria:

\begin{itemize}

\item We include $f\sigma_8(z)$ measurements from uncorrelated redshift bins when they are based on the same cosmological tracer. Data from potentially correlated redshift bins are also considered, but they correspond to different cosmological tracers.

\item Only direct measurements of $f\sigma_8(z)$ are included.

\item For surveys with multiple $f\sigma_8(z)$ measurements across different data releases, only the most recent measurements are considered.

\end{itemize}

The compilation above may include fewer data points than typically assumed in recent literature, but it ensures that our results remain statistically consistent by adhering to a rigorous selection criteria for RSD data.

\item \textit{Cosmic Microwave Background} (\textbf{CMB}): 
Temperature and polarization anisotropy measurements of the CMB power spectra (as well as their cross-spectra) from the Planck 2018 legacy data release (PR3). In particular we use: 
\begin{itemize}
\item the high-$\ell$ \texttt{Plik} likelihood~\cite{Planck:2018_3} for the TT spectrum in the multipole range $30 \leq \ell \leq 2508$ as well as for the TE and EE spectra at $30 \leq \ell \leq 1996$; 
\item the low-$\ell$ \texttt{commander} likelihood~\cite{Planck:2018_3} for the TT spectrum in the multipole range $2 \leq \ell \leq 29$;
\item the low-$\ell$ \texttt{SimAll} likelihood~\cite{Planck:2018_3} for the EE spectrum at $2 \leq \ell \leq 29$.
\item the \texttt{Plik} CMB Planck lensing measurements~\cite{Planck:2018lbu}, reconstructed from the temperature 4-point correlation function.
\end{itemize}

\item \textit{Baryon Acoustic Oscillations} (\textbf{DESI}):
DESI BAO measurements obtained from observations of galaxies and quasars~\cite{DESI:2024uvr}, and Lyman-$\alpha$~\cite{DESI:2024lzq} tracers, as summarized in Table I of Ref.~\cite{DESI:2024mwx}. These measurements consist of both isotropic and anisotropic BAO data in the redshift range $0.1 < z < 4.2$ and are divided into seven redshift bins. The isotropic BAO measurements are represented as $D_{\mathrm{V}}(z)/r_{\mathrm{d}}$, where $D_{\mathrm{V}}$ denotes the angle-averaged distance, normalized to the (comoving) sound horizon at the drag epoch. The anisotropic BAO measurements include $D_{\mathrm{M}}(z)/r_{\mathrm{d}}$ and $D_{\mathrm{H}}(z)/r_{\mathrm{d}}$, where $D_{\mathrm{M}}$ is the comoving angular diameter distance and $D_{\mathrm{H}}$ is the Hubble horizon. Additionally, the correlation between the measurements of $D_{\mathrm{M}}/r_{\mathrm{d}}$ and $D_{\mathrm{V}}/r_{\mathrm{d}}$ is also taken into account.

\item \textit{Type Ia Supernovae} (\textbf{PP}): Type Ia Supernovae (SNe Ia) are widely used as standard candles due to their relatively uniform absolute luminosity~\cite{riess1998observational}. We incorporated SNe Ia distance modulus measurements from the PantheonPlus sample, which consists of 1550 supernovae spanning a redshift range from $0.01$ to $2.26$~\cite{pantheonplus}.

\item \textit{Cosmic Chronometers} (\textbf{CC}): 
measurements of the expansion rate $H(z)$ derived from the relative ages of massive, early-time, passively-evolving galaxies, known as Cosmic Chronometers 
~\cite{Jimenez:2001gg}. In our analyses, we conservatively used only a compilation of 15 CC measurements in the redshift range from $z\sim0.179$ to $z\sim1.965$~\cite{Moresco:2012by,Moresco:2015cya,Moresco:2016mzx}, accounting for all non-diagonal terms in the covariance matrix and systematic contributions.

\end{itemize}

As a final remark, we emphasize that in all analyses without CMB data, we use state-of-the-art assumptions on Big Bang Nucleosynthesis (BBN). Specifically, the BBN data consist of measurements of the primordial abundances of helium, $Y_P$, from~\cite{Aver_2015}, and the deuterium measurement, $y_{\rm DP} = 10^5 n_D / n_H$, obtained in~\cite{Cooke_2018}. This BBN likelihood is sensitive to the constraints on the physical baryon density $\omega_{\rm b}$ and the effective number of neutrino species $N_{\rm eff}$, where in this work we keep fixed to $N_{\rm eff} = 3.044$.

\begin{table*}[htpb!]
\begin{center}
\caption{Marginalized constraints on the interaction parameter $\xi$ and other derived parameters of the IDE model with $\xi < 0$, based on different datasets: RSD, DESI+PP, RSD+DESI+PP, and RSD+DESI+PP+CC. For each dataset, the mean values with $68\%$ and $95\%$ CL are provided. Additionally, the table shows $\Delta \chi^2_{\text{min}} = \chi^2_{\text{min (IDE)}} - \chi^2_{\text{min ($\Lambda$CDM)}}$, where negative values indicate a better fit of the IDE model compared to the $\Lambda$CDM model.}
\label{tab.rsd_xi_negative}
\renewcommand{\arraystretch}{1.5}
\resizebox{\textwidth}{!}{
\begin{tabular}{lcccc} \hline
\textbf{Parameter} & \textbf{RSD} & \textbf{DESI+PP} & \textbf{DESI+PP+RSD} & \textbf{DESI+PP+RSD+CC} \\ \hline \hline
$\xi$ & $\xi > -0.0250$ ($\xi > -0.0440$) & $\xi > -0.0729$ ($\xi > -0.177$) & $\xi > -0.0200$ ($\xi > -0.0430$) & $\xi > -0.0197$ ($\xi > -0.0422$)  \\
$\Omega_\mathrm{m}$ & $0.253^{+0.053}_{-0.074}$ ($0.25^{+0.13}_{-0.12}$) & $0.297^{+0.016}_{-0.013}$ ($0.297^{+0.030}_{-0.031}$) & $0.304\pm 0.012$ ($0.304^{+0.024}_{-0.023}$) & $0.304\pm 0.012$ ($0.304^{+0.023}_{-0.022}$)  \\
$S_8$ & $0.738\pm 0.068$ ($0.74\pm 0.14$) & $-$ & $0.775\pm 0.027$ ($0.775^{+0.052}_{-0.053}$) & $0.775\pm 0.026$ ($0.775^{+0.052}_{-0.052}$)  \\
$H_0 \mathrm{(km/s/Mpc)}$ & $-$ & $70.17^{+0.82}_{-1.0}$ ($70.2^{+1.9}_{-1.8}$) & $69.73\pm 0.78$ ($69.7^{+1.5}_{-1.5}$) & $69.67\pm 0.75$ ($69.7^{+1.5}_{-1.4}$)  \\
\hline
$\Delta \chi^{2}_{\mathrm{min}}$ & $0.031$ & $-0.0009$ & $-0.042$ & $0.048$ \\
\hline \hline \end{tabular}}
\end{center}
\end{table*}

\begin{figure*}[htpb!]
    \centering
    \includegraphics[width=0.65\textwidth]{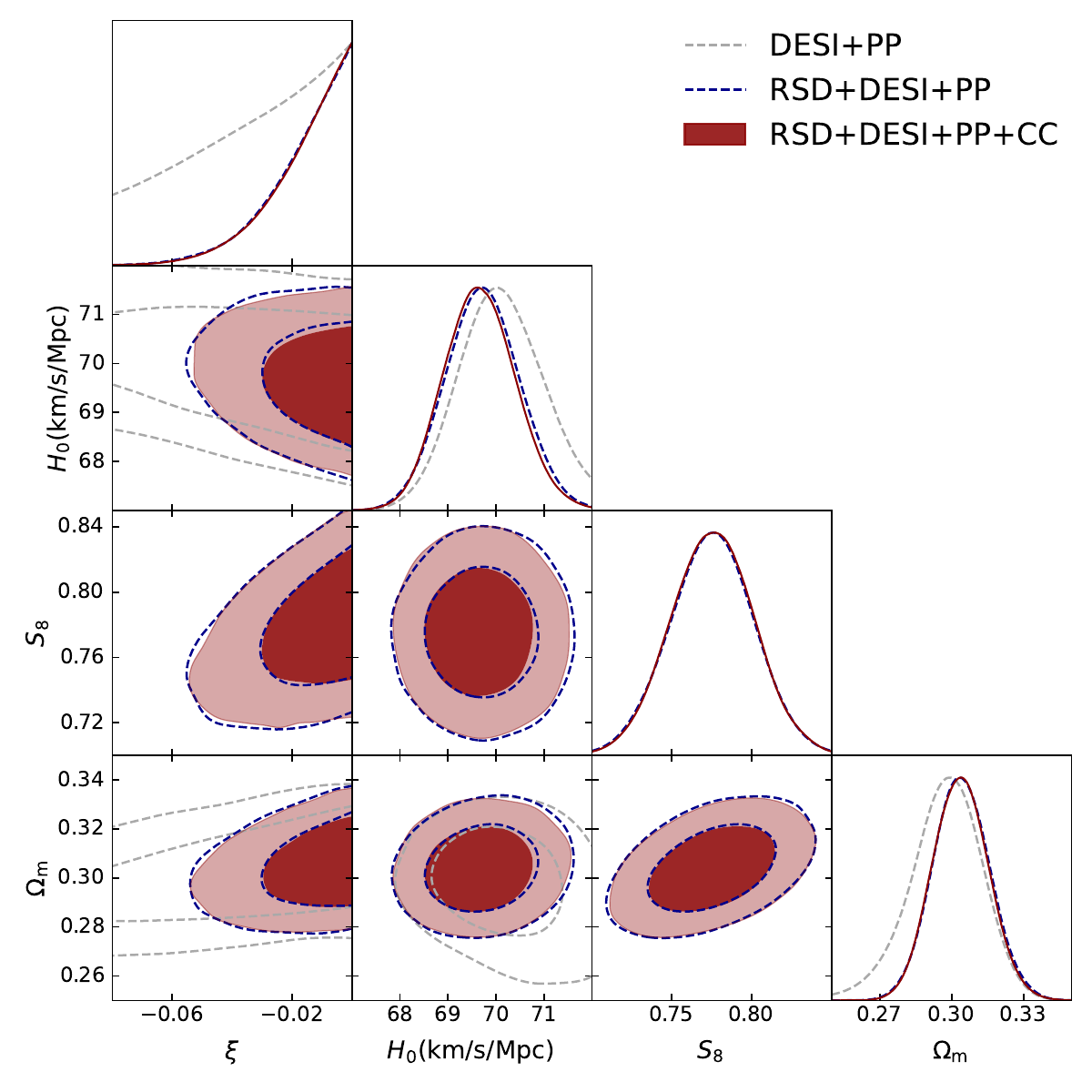} \,\,\,
    \caption{Marginalized posterior distributions and 68\% and 95\% CL contours for the parameters $\Omega_{\rm m}$, $\xi$, $H_{0}$, and $S_{8}$ in the IDE model with $\xi < 0$. The results are shown for different combinations of datasets, as indicated in the legend.}
    \label{fig:rsd_xi_negative}
\end{figure*}

\begin{table*}
\begin{center}
\caption{Marginalized constraints on the parameters $\xi$ and some derived parameters of the IDE model with $\xi > 0$, including mean values with 68\% and 95\% CL. The results are shown for different datasets: RSD, DESI+PP, RSD+DESI+PP, and RSD+DESI+PP+CC. The table also includes $\Delta \chi^2_{\text{min}} = \chi^2_{\text{min (IDE)}} - \chi^2_{\text{min ($\Lambda$CDM)}}$, where negative values indicate a better fit for the IDE model compared to the $\Lambda$CDM model.}
\label{tab.rsd_xi_positive}
\renewcommand{\arraystretch}{1.5}
\resizebox{\textwidth}{!}{
\begin{tabular}{lcccc} \hline
\textbf{Parameter} & \textbf{RSD} & \textbf{DESI+PP} & \textbf{DESI+PP+RSD} & \textbf{DESI+PP+RSD+CC} \\ \hline \hline
$\xi$ & $\xi < 0.575$ ($\xi < 0.926$) & $0.26^{+0.12}_{-0.15}$ ($0.26^{+0.23}_{-0.25}$) & $\xi < 0.0396$ ($\xi < 0.0717$) & $\xi < 0.0396$ ($\xi < 0.0716$)  \\
$\Omega_\mathrm{m}$ & $0.582^{+0.18}_{-0.083}$ ($0.58^{+0.20}_{-0.28}$) & $0.356^{+0.024}_{-0.029}$ ($0.356^{+0.051}_{-0.050}$) & $0.317\pm 0.013$ ($0.317^{+0.026}_{-0.024}$) & $0.316^{+0.012}_{-0.013}$ ($0.316^{+0.026}_{-0.024}$)  \\
$S_8$ & $1.08^{+0.19}_{-0.11}$ ($1.08^{+0.24}_{-0.28}$) & $-$ & $0.812^{+0.027}_{-0.031}$ ($0.812^{+0.060}_{-0.054}$) & $0.812^{+0.028}_{-0.032}$ ($0.812^{+0.061}_{-0.057}$)  \\
$H_0 \mathrm{(km/s/Mpc)}$ & $-$ & $67.0^{+1.6}_{-1.4}$ ($67.0^{+2.8}_{-3.0}$) & $69.37\pm 0.78$ ($69.4^{+1.6}_{-1.5}$) & $69.31\pm 0.77$ ($69.3^{+1.5}_{-1.5}$)  \\
\hline
$\Delta \chi^{2}_{\mathrm{min}}$ & $-0.15$ & $-3.45$  & $-0.37$ & $-0.26$  \\
\hline \hline \end{tabular}}
\end{center}
\end{table*}

\begin{figure*}[htpb!]
    \centering
    \includegraphics[width=0.65\textwidth]{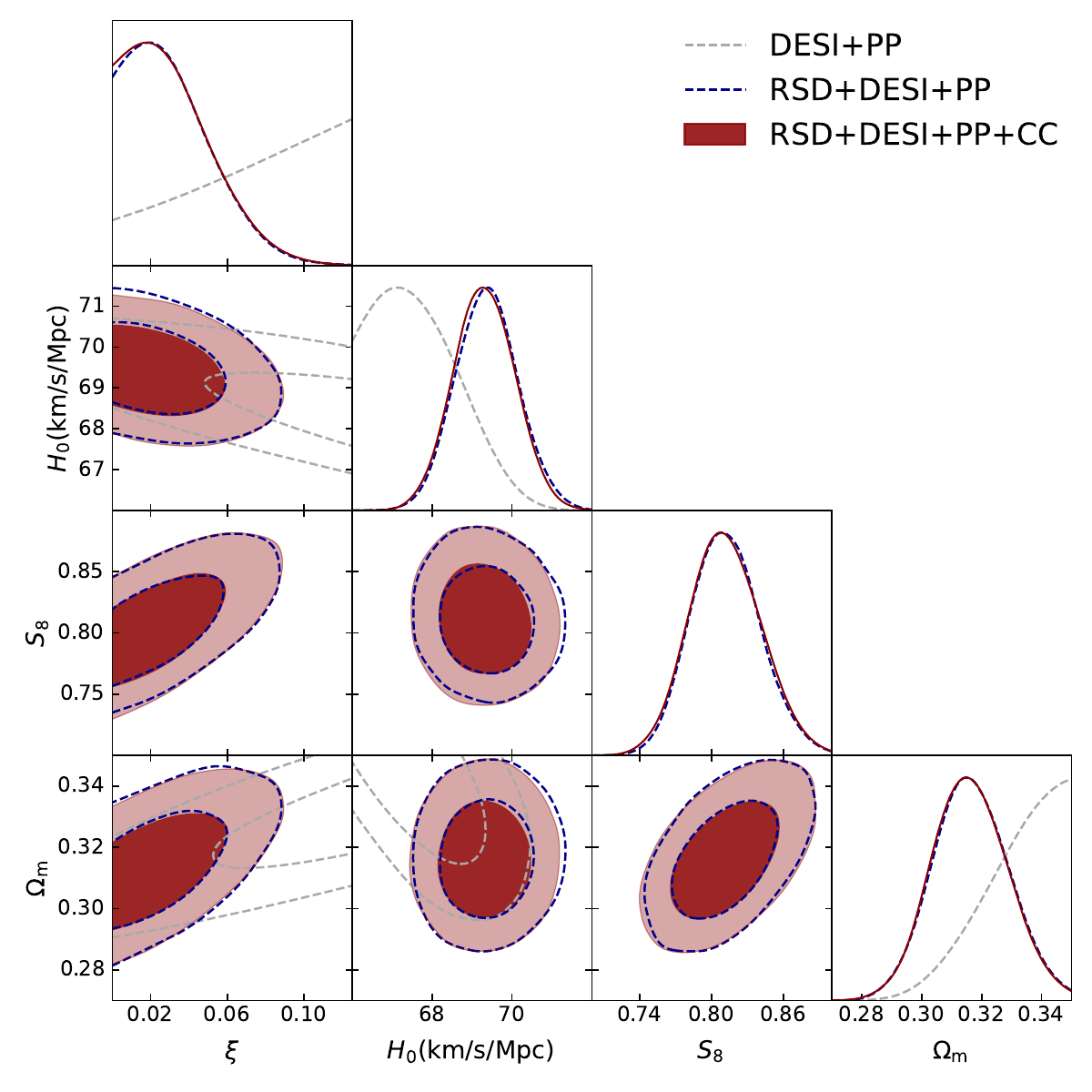} \,\,\,
     \caption{Marginalized posterior distributions and contours (68\% and 95\% CL) for the parameters $\Omega_{\rm m}$, $\xi$, $H_{0}$, and $S_{8}$ in the IDE model with $\xi > 0$. The results are shown for various combinations of data sets as indicated in the legend.}
    \label{fig:rsd_xi_positive}
\end{figure*}

\section{Results and Discussions}
\label{results}

In this section, we will discuss and present our main results. 
We distinguish between the results obtained with and without the inclusion of CMB data. In each case, we present the results obtained for the parameters of interest in the model -- primarily the coupling parameter $\xi$ and its correlations with other cosmological parameters such as $H_0$ and $S_8$ related to cosmic tensions. Additionally, we will present a detailed discussion of the degree of tension between different datasets. When discussing the degree of tension between two datasets, we will employ the quadratic estimator proposed in Ref.~\cite{Addison:2015wyg}. The estimator is given by:
\begin{equation}
    \chi^2 = \left(\mathbf{x}_i - \mathbf{x}_j\right)^\mathrm{T} \left(\mathcal{C}_i + \mathcal{C}_j\right)^{-1} \left(\mathbf{x}_i - \mathbf{x}_j\right),
    \label{N-tension}
\end{equation}
where $ \mathbf{x}_i $ and $ \mathbf{x}_j $ are the vectors containing the mean values for the cosmological parameters inferred from datasets $ i $ and $ j $, respectively. Similarly, $ \mathcal{C}_i $ and $ \mathcal{C}_j $ are the corresponding covariance matrices. This estimator allows us to quantify the level of agreement or tension between the datasets, providing a clear statistical measure of how well the datasets align with each other. This method is particularly useful for identifying discrepancies that could indicate systematic errors or the need for new theoretical models.


\begin{table*}[htpb!]
\begin{center}
\caption{Marginalized constraints, mean values with $68\%$ ($95\%$) CL, on the free and some derived parameters of the IDE ($\xi > 0$) model from the CMB dataset and its combinations with the RSD, DESI, PP, and CC datasets. For all datasets, we present $\Delta \chi^2_{\text{min}} = \chi^2_{\text{min (IDE)}} - \chi^2_{\text{min ($\Lambda$CDM)}}$, where negative values signify a superior fit and an inclination towards the IDE model over the $\Lambda$CDM model.}
\label{tab_results_positiveCMB}
\renewcommand{\arraystretch}{1.75}
\resizebox{\textwidth}{!}{
\large
\begin{tabular}{lcccc} 
\hline
\textbf{Parameter} & \textbf{CMB} & \textbf{CMB+RSD} & \textbf{CMB+RSD+DESI} & \textbf{CMB+RSD+DESI+PP+CC}\\ 
\hline\hline
$\Omega_{\rm b} h^2$ & $0.02238\pm 0.00014$ ($0.02238^{+0.00028}_{-0.00028}$) & $0.002237\pm 0.00015$ ($0.002237^{+0.00029}_{-0.00028}$) &
$0.002253\pm 0.00013$ ($0.002253^{+0.00029}_{-0.00027}$) & $0.002250\pm 0.00013$ ($0.002250^{+0.00027}_{-0.00026}$)  \\

$\Omega_{\rm c} h^2 $ & $0.1350^{+0.0072}_{-0.013}$ ($0.135^{+0.020}_{-0.016}$) & $0.1227^{+0.0018}_{-0.0024}$ ($0.1227^{+0.0044}_{-0.0040}$) & $0.1199^{+0.0013}_{-0.0019}$ ($0.1199^{+0.0033}_{-0.0029}$) & $0.1207^{+0.0014}_{-0.0019}$ ($0.1207^{+0.0033}_{-0.0030}$)  \\

$100\theta_\mathrm{s}$ & $1.04187\pm 0.00029$ ($1.04187^{+0.00056}_{-0.00056}$) & $1.04189\pm 0.00030$ ($1.04189^{+0.00060}_{-0.00055}$) & 
$1.04206\pm 0.00028$ ($1.04206^{+0.00055}_{-0.00057}$) & 
$1.04204\pm 0.00028$ ($1.04204^{+0.00056}_{-0.00053}$) \\

$\tau_\mathrm{reio}$ & $0.0544\pm 0.0076$ ($0.054^{+0.016}_{-0.014}$) & $0.0543^{+0.0065}_{-0.0077}$ ($0.054^{+0.015}_{-0.013}$) & $0.0603^{+0.0069}_{-0.0082}$ ($0.060^{+0.016}_{-0.014}$) & 
$0.0590\pm 0.0071$ ($0.059^{+0.015}_{-0.013}$)   \\

$n_\mathrm{s}$ & $0.9651\pm 0.0040$ ($0.9651^{+0.0080}_{-0.0080}$) & $0.9655\pm 0.0041$ ($0.9655^{+0.0082}_{-0.0078}$) &
$0.9705\pm 0.0037$ ($0.9705^{+0.0075}_{-0.0073}$) & 
$0.9695\pm 0.0035$ ($0.9695^{+0.0071}_{-0.0067}$)  \\

$\log(10^{10} A_\mathrm{s})$ & $3.046\pm 0.015$ ($3.046^{+0.032}_{-0.028}$) & $3.045\pm 0.014$ ($3.045^{+0.029}_{-0.027}$) & 
$3.053^{+0.014}_{-0.016}$ ($3.053^{+0.031}_{-0.029}$) & 
$3.051\pm 0.014$ ($3.051^{+0.028}_{-0.026}$) \\

$\xi$ & $< 0.205$ ($< 0.388$) & $0.0250^{+0.0079}_{-0.023}$ ($< 0.0545$) & $< 0.0214 $ ($< 0.0405$)  & $< 0.0259$ ($< 0.0449$)  \\\hline

$H_0$ (km/s/Mpc) & $65.8^{+1.5}_{-0.99}$ ($65.8^{+2.1}_{-2.6}$) & $67.11\pm 0.58$ ($67.1^{+1.2}_{-1.1}$) &
$68.13\pm 0.43$ ($68.13^{+0.83}_{-0.82}$) & $67.92\pm 0.41$ ($67.92^{+0.81}_{-0.77}$)  \\

$\Omega_{\rm m }$ & $0.366^{+0.025}_{-0.048}$ ($0.366^{+0.076}_{-0.059}$) & $0.3238^{+0.0088}_{-0.010}$ ($0.324^{+0.020}_{-0.018}$) &
$0.3084^{+0.0060}_{-0.0072}$ ($0.308^{+0.014}_{-0.012}$) &
$0.3118^{+0.0062}_{-0.0071}$ ($0.312 \pm 0.013$)  \\

$S_8$ & $0.787\pm 0.024$ ($0.787^{+0.044}_{-0.046}$) & $0.813\pm 0.012$ ($0.813^{+0.024}_{-0.024}$) & $0.796\pm 0.010$ ($0.796\pm 0.019$) & $0.798\pm 0.010$ ($0.798^{+0.019}_{-0.019}$)  \\ 

\hline
$\Delta \chi^{2}_{\mathrm{min}}$ & $0.9$ & $-1.98$ & $-1.13$ & $-0.76$  \\
\hline \hline
\end{tabular}
}
\end{center}
\end{table*}

\begin{figure*}[t]
    \centering
    \includegraphics[scale=0.5]{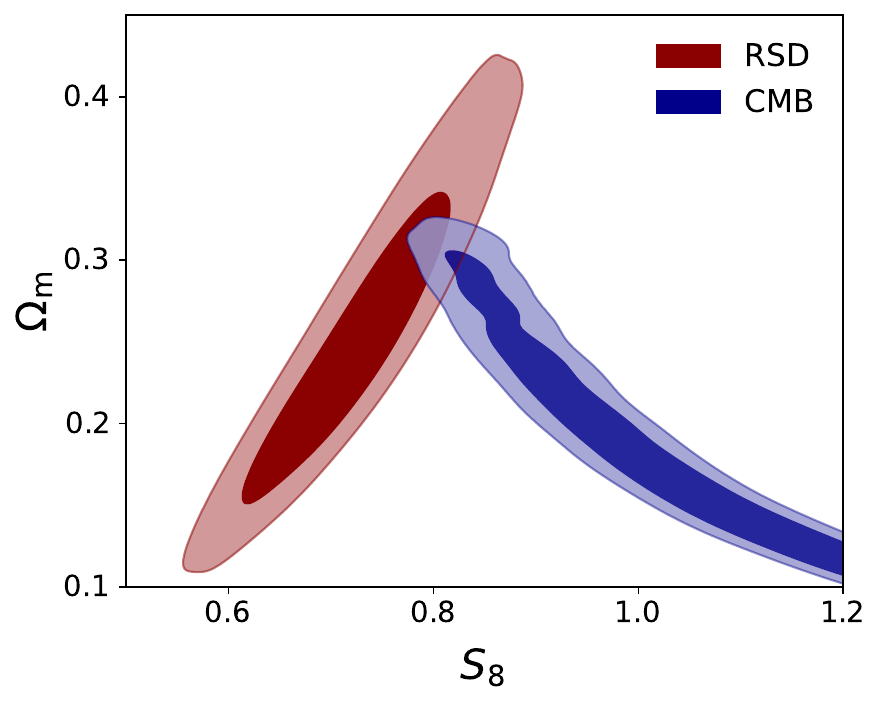}
    \includegraphics[scale=0.5]{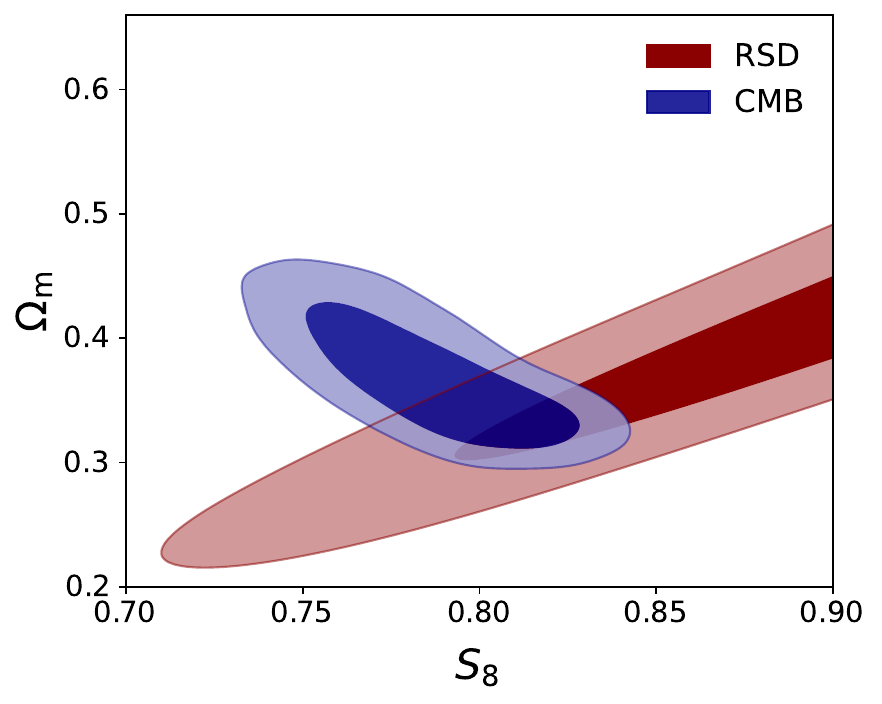}
    \caption{Left panel: Marginalized posterior contours (68\% and 95\% CL) in the $S_{8}$-$\Omega_{\rm m}$ plane for the IDE model with $\xi < 0$. Right panel: Same as the left panel, but for the IDE model with $\xi > 0$.}
    \label{fig:s8_xi}
\end{figure*}

\begin{figure*}[htpb!]
    \centering
    \includegraphics[width=0.65\textwidth]{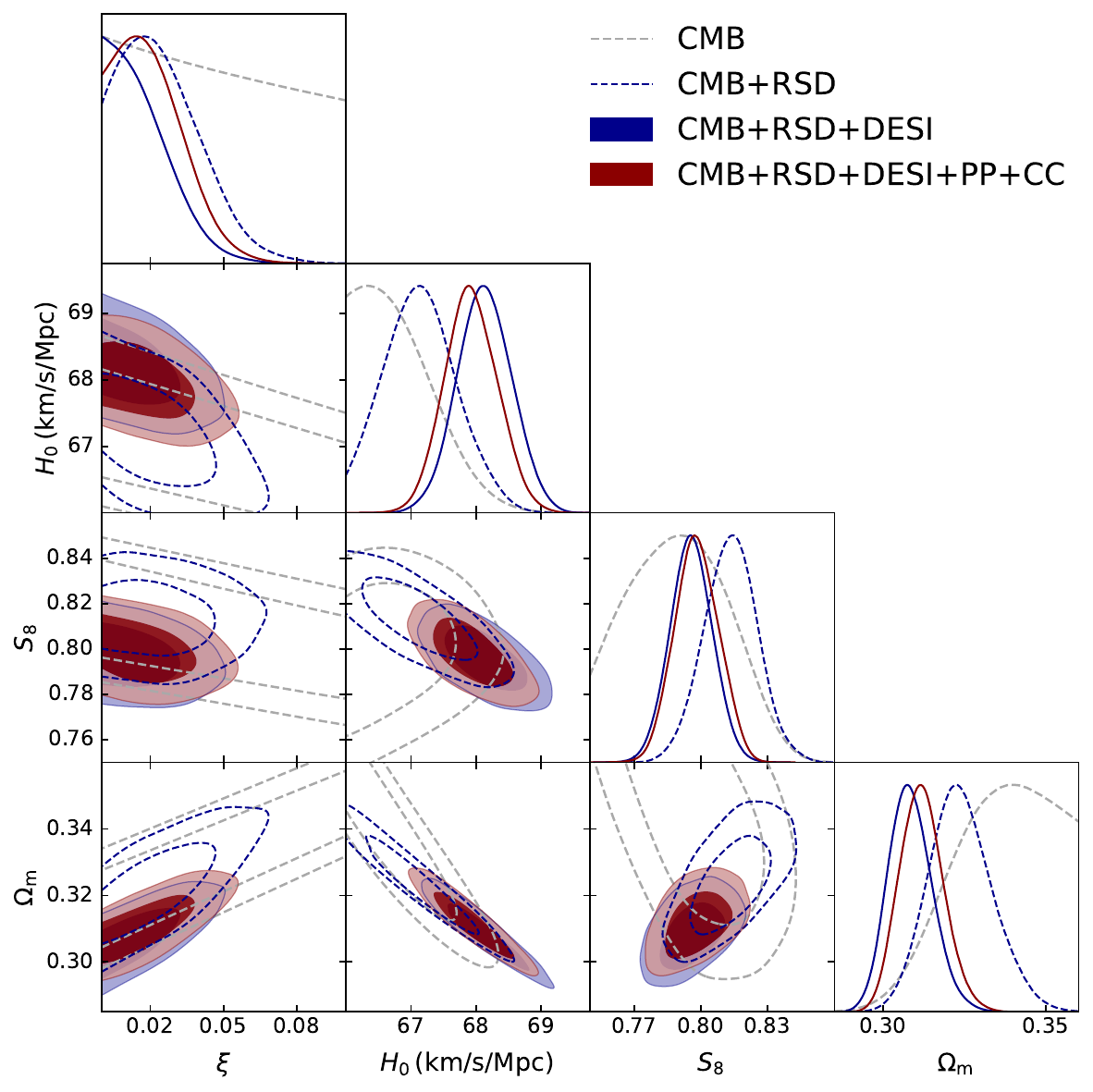}
    \caption{Marginalized posterior distributions and contours (68\% and 95\% CL) for $\Omega_{\rm m}$, $\xi$, $H_{0}$, and $S_{8}$ parameters of the IDE ($\xi > 0$) model for various data combinations listed in the legend.}
    \label{fig:cmb_xi_positive}
\end{figure*}

\subsection{Constraints without CMB data}

We begin by exploring the observational constraints 
obtained \textit{without} the inclusion of CMB data. As established in prior studies (see, e.g., Ref.~\cite{Zhai_2023}), the model under investigation fits very well with various current CMB experiments, all preferring the same value for the interaction. Therefore, in this section, we emphasize the constraints derived from the RSD data. It is also important to note that we are using the most 
recent BAO samples available in the literature, obtained through the DESI survey.

In Table~\ref{tab.rsd_xi_negative}, we summarize the results for the case where $\xi < 0$. As anticipated, RSD data alone do not possess the capability to robustly restrict the entire parameter space of the model. However, several noteworthy points emerge. As expected, the RSD data are not sensitive to the Hubble constant, which prevents us from obtaining precise constraints on $H_0$ from RSD data alone. The constraints on $\Omega_{\rm m}$ and $S_8$ show significantly lower values compared to those predicted by the $\Lambda$CDM model, albeit with large error bars. However, the most interesting observation is that the coupling parameter $\xi$ is well-constrained using RSD data alone. 
Specifically, we find $\xi > -0.0440$ at 95\% CL, indicating that RSD can impose strong bounds on the total amount of energy-momentum that can be transferred from DM to DE. Notably, in this case, RSD data improve the constraints on the coupling parameter by an order of magnitude compared to CMB data alone (see CMB-only constraints in~\cite{Zhai_2023}).

In the second column of Table~\ref{tab.rsd_xi_negative}, we present the joint analysis with DESI and PP data, focusing only on geometrical constraints, excluding RSD predictions. Since this dataset does not provide constraints on $\sigma_8 - S_8$, we only have information on the other parameters within the common baseline. As explored in previous works, there is a significant degeneracy in the $\xi - H_0$ plane, as also illustrated in Figure~\ref{fig:rsd_xi_negative}. The joint analysis of DESI+PP data does not indicate any evidence for the coupling parameter and provides $\xi > -0.177$ at 95\% CL.  However, they prefer higher values of $H_0$, bringing the tension with the SH0ES down to $2.2\sigma$.

As a next step, we perform a joint analysis of RSD+DESI+PP data. We obtain tight constraints on the entire parameter space of the model. Specifically, the coupling parameter is constrained to $\xi > -0.0430$ at 95\% CL. The parameter $S_8$ is robustly constrained to $S_8 = 0.775 \pm 0.027$ at 68\% CL, indicating potentially low values of $S_8$ from the combination of RSD+DESI+PP data. However, we do not observe any evidence supporting a non-zero coupling parameter, as it is tightly constrained by the joint analysis. As quantified in Table~\ref{tab.rsd_xi_negative} and Figure~\ref{fig:rsd_xi_negative}, the addition of CC data does not result in significant statistical improvements. Therefore, RSD+DESI+PP alone is robust enough to constrain the entire parameter space of this model. These analyses clearly demonstrate that RSD data improve the constraints significantly on this class of models.

Now, we turn our attention to the scenario where $\xi > 0$. We summarize the results for this case in Table~\ref{tab.rsd_xi_positive}. 
When considering RSD data, we observe behavior opposite to the $\xi < 0$ case. This reflects the opposite direction in which the energy-momentum is transferred within the dark sector (here from DE to DM), introducing opposite correlations between the coupling $\xi$ and the other baseline parameters. Specifically, the constraints on the parameters $\Omega_{\rm m}$, $H_0$, and $S_8$ move in the opposite direction. With $\xi > 0$, RSD data tend to predict higher values for both $\Omega_{\rm m}$ and $S_8$, although the error bars remain large since RSD data alone lack strong constraining power. Overall, when $\xi > 0$, RSD data on their own do not provide robust constraints on the free parameters of the model, including $\xi$. 

In the second column of Table~\ref{tab.rsd_xi_positive}, we present the joint analysis of DESI and PP data. Here, we would like to highlight an important result: considering only DESI+PP data, we find $\xi = 0.26^{+0.12}_{-0.15}$ at 68\% CL, suggesting a preference for non-vanishing $\xi > 0$ at approximately $1.8\sigma$. 
This result can be understood as follows: using the continuity equations, after simple algebraic manipulations, the effective equation of state parameter of DE in the model is given by~\cite{Kumar:2021eev} 
\begin{equation}
w_{\rm eff} = -1 + \frac{\xi}{3}.
\end{equation}
Thus, for $\xi > 0$ ($\xi < 0$), the model predicts quintessence-type (phantom-type) dynamics. Therefore, the joint analysis of DESI+PP supports quintessence-type effective dynamics, in line with the DESI results~\cite{DESI:2024mwx}.

Next, we pursue the analysis with the combined data from RSD+DESI+PP. In this case, we find $\xi < 0.0717$ at 95\% CL. This joint analysis robustly constrains the full parameter space of the model. We obtain $H_0 = 69.37 \pm 0.78$ km/s/Mpc, which is in tension with local measurements by the SH0ES team. In contrast to the $\xi < 0$ case, we observe a slightly higher value for $S_8$. This is due to the strong correlation between $\xi$ and $\sigma_8$, which acts in opposite directions depending on the sign of $\xi$.

Figure~\ref{fig:rsd_xi_positive} shows the one-dimensional and two-dimensional marginalized posterior distributions (68\% and 95\% CL) for the parameters of interest in the model, based on our statistical analyses. 

It is evident from these contours that the inclusion of RSD data significantly enhances the constraining power on the parameters, leading to a more precise determination of their values and a reduction in uncertainties. Notably, the linear perturbation effects captured by RSD measurements impose stronger upper bounds on the coupling parameter $\xi$ in this context. Additionally, the inclusion of CC data does not impact the results for $\xi > 0$, further reinforcing that RSD+PP+DESI alone provides robust constraints on the model parameters.

\subsection{Constraints with CMB data}
In this section, we will discuss observational constraints, with a particular focus on CMB data. We begin by examining the scenario where $\xi < 0$. The first point of investigation is the internal consistency between the CMB data and RSD measurements. By using Eq.~\eqref{N-tension}, we find a tension of 2.2$\sigma$ between the CMB and RSD datasets 
which is primarily driven by discrepancies in the values of $S_8$ and $\Omega_{\rm m}$ inferred from these two probes; see also the left panel of Figure~\ref{fig:s8_xi}.
Given the significant magnitude of this disagreement, we opt not to combine the CMB and RSD datasets when $\xi < 0$. 
Conversely, in the scenario where $\xi > 0$, we observe only a minor tension between the CMB and RSD datasets. The probability contours for $S_8$ and $\Omega_{\rm m}$ from the two probes are found to be in reasonable agreement, as clearly illustrated in the right panel of Figure~\ref{fig:s8_xi}. We find that the tension in this case is less than 2$\sigma$. Therefore, in this case, it is legitimate to combine CMB and RSD data with each other and with the other datasets used in this study. 

To avoid combining datasets with significant tension that could possibly bias results for crucial parameters, such as the coupling parameter quantifying energy-momentum transfer within the dark sector (which, as our CMB-free analysis shows, is highly correlated with parameters exhibiting the greatest tension), we will focus on the case $\xi > 0$. In Table \ref{tab_results_positiveCMB}, we summarize our statistical results, showing the constraints at 68\% and 95\% CL for our baseline parameters of interest. When analyzing CMB data only, 
we find an upper limit for the coupling parameter $\xi < 0.388$ at 95\% CL. As expected, the strong correlation between $\xi$ and the other parameters (see Figure~\ref{fig:cmb_xi_positive}) drives $H_0$ and $S_8$ towards lower values compared to the case where $\xi < 0$. As a result, while IDE models with energy-momentum transfer from DE to DM do not alleviate the Hubble tension, they can significantly help address (or at least relax) the tension in $S_8$. The predicted $S_8$ value is low enough to be fully compatible with the RSD-only analysis discussed in the previous section, as well as with cosmic shear measurements. 
Again, this is clearly illustrated in the right panel of Figure~\ref{fig:s8_xi}, where the probability contours for $S_8$ and $\Omega_m$ inferred from the two probes overlap well within the 95\% CL region of the parameter space. This comprehensive analysis underscores the importance of considering the correlations between parameters and highlights how certain scenarios can effectively reconcile discrepancies between different observational datasets. 

In the joint CMB+RSD analysis, we observe evidence for an energy transfer from DE to DM at 68\% CL, although this effect is consistent with zero at 95\% CL. Specifically, we find that $\xi < 0.0545$ at 95\% CL in this joint analysis. This result is noteworthy, clearly indicating that the combination of evolution-dependent observables from linear perturbations of matter with CMB leads to an improvement in the upper limits when compared to BAO and SNe Ia data.

As a next step, we explore different combinations involving DESI and PP data. The inclusion of these observations not only improves the precision of the constraints but also results in slightly lower $S_8$ values. For the CMB+RSD+DESI combination, we find an upper limit of $\xi < 0.0405$. In the case of the CMB+RSD+DESI+PP+CC combination, the upper limit shifts to $\xi < 0.0449$. The slight increase in the upper limit of $\xi$ is driven by the influence of SNIa data, which tends to pull $\xi$ towards more positive values. It is evident that RSD predictions play a crucial role in constraining the parameter space for this class of models, resulting in such tight upper limits. We refer to Figure~\ref{fig:cmb_xi_positive} for the one- and two-dimensional marginalized probability distributions of the parameters $\Omega_{\rm m}$, $\xi$, $H_0$, and $S_8$ obtained from all these data combinations.

\begin{figure*}[t]
   \includegraphics[scale=0.32]{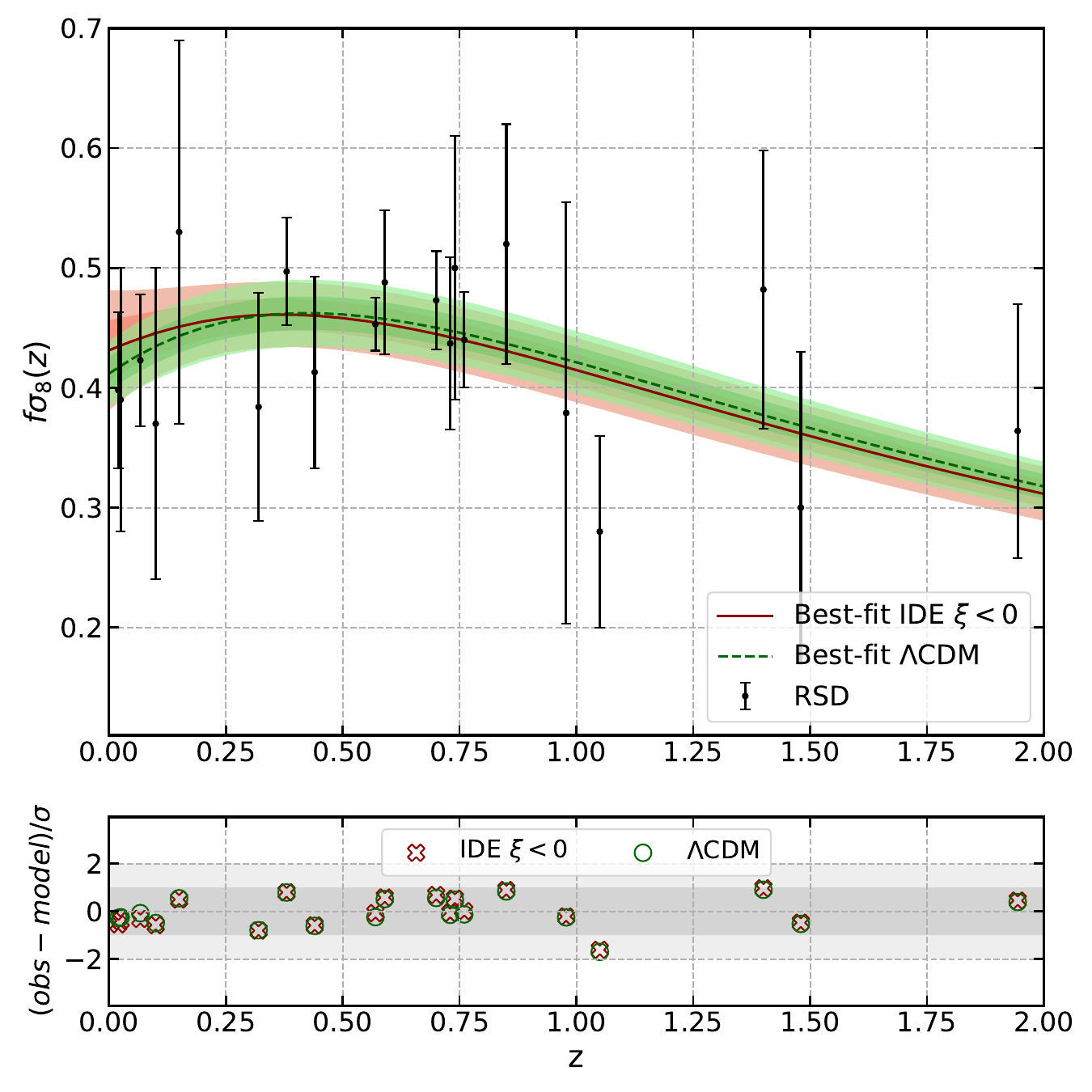} \,\,\,\,
   \includegraphics[scale=0.32]{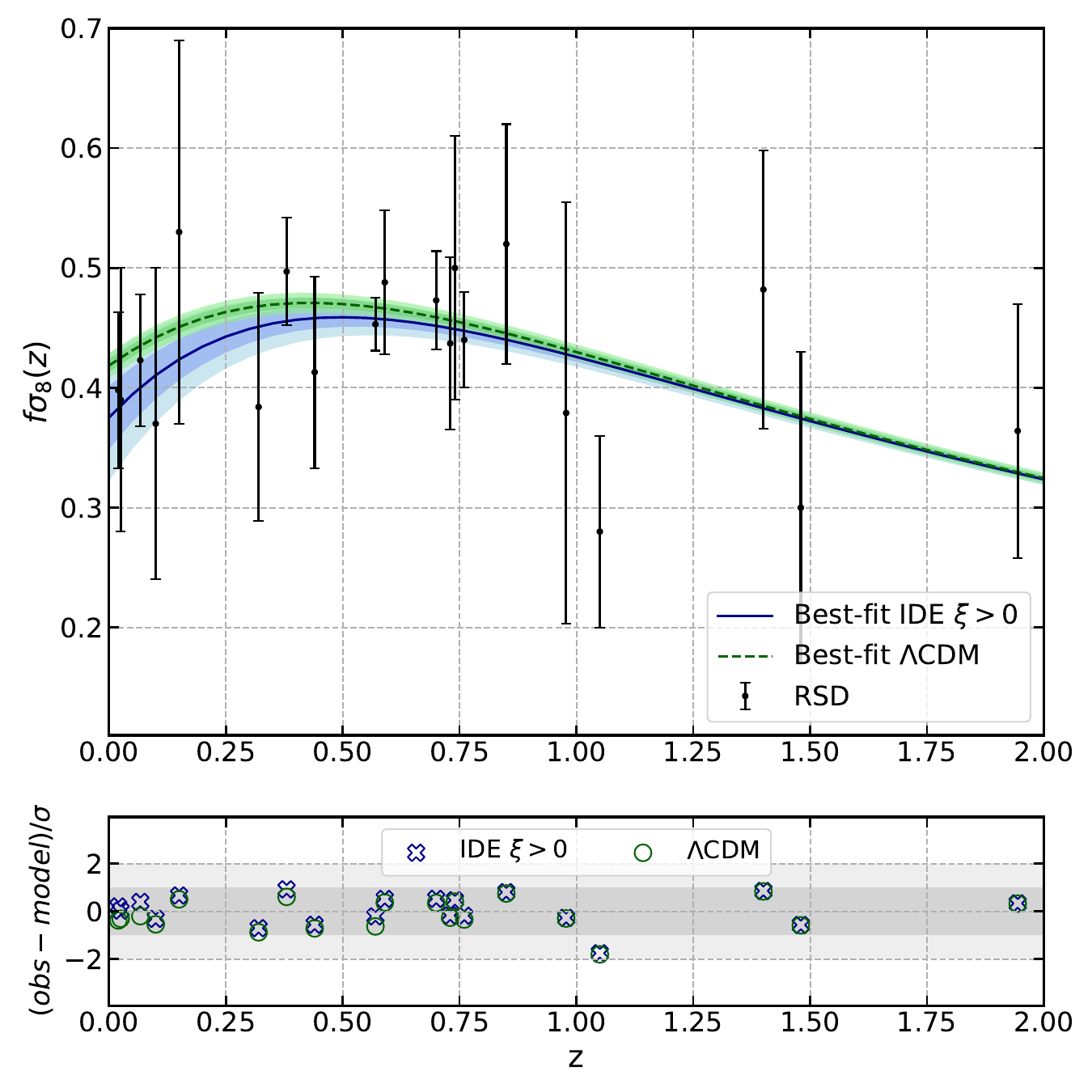} 
   \caption{Left panel: Statistical reconstruction of the theoretical prediction \( f\sigma_8(z) \) at 1$\sigma$ and 2$\sigma$ confidence levels for the $\Lambda$CDM and IDE models with $\xi < 0$ through the joint analysis of RSD+DESI+PP+CC, compared to RSD measurements. Right panel: Same as the left panel, but assuming the joint analysis of CMB+RSD+DESI+PP+CC for the $\Lambda$CDM and IDE models with $\xi > 0$. The lower panel of both figures shows the difference between the model prediction and each RSD measurement, normalized by the observational uncertainties.}
   \label{fig:fs8_CMB_reconstruction}
\end{figure*}

At this stage, it is useful to compare our results for $S_8$ with the current weak lensing and galaxy clustering measurements predicted by surveys such as DES \cite{DES_2021} and KiDS-1000 \cite{KiDS:2020suj}. From our analysis based on CMB data alone, we find $S_8 = 0.787 \pm 0.024$, a value that is slightly low and consistent with the cosmic shear measurements from DES and KiDS-1000. When considering our joint analysis, which incorporates CMB, RSD, and DESI data, we find $S_8 = 0.796\pm 0.010$, which exhibits a tension of approximately $1.5\sigma$ with KiDS-1000, and a similar degree of tension with the DES results. Since these models modify the late-time growth of cosmic structures, a more quantitative comparison requires re-analyzing the weak lensing and galaxy clustering data from these surveys, as the constraints from the cosmic shear samples depend on the specific cosmological model. In future work, we will explore this issue in more detail to address the model-dependence of these comparisons.

Overall, our comprehensive analysis reinforces the importance of RSD measurements in providing strong observational constraints on IDE, regardless of the sign of the interaction between the dark components.
To acquire a better understanding of the origin of this constrains, in Figure~\ref{fig:fs8_CMB_reconstruction} we present a statistical reconstruction of the theoretical predictions for $f\sigma_8(z)$ at 1$\sigma$ and 2$\sigma$ CL, as obtained within the $\Lambda$CDM and IDE models. The lower panel of the figure shows the difference between the model prediction and each RSD measurement, normalized by the observational uncertainties. Focusing on the RSD+DESI+PP+CC dataset, the IDE model with $\xi < 0$ shows statistical consistency with the $\Lambda$CDM model across the entire $z$-range -- see also the left panel of Figure~\ref{fig:fs8_CMB_reconstruction}. Conversely, the combination of CMB+RSD+DESI+PP+CC data reveals that the presence of a DE-DM interaction with $\xi > 0$ can significantly impacts the growth factor of structures at $z < 0.5$, resulting in a greater suppression in the late universe, as clearly seen in the right panel of Figure~\ref{fig:fs8_CMB_reconstruction}.

In all the tests we conducted, we observed that measurements at low redshift values (for $z < 0.5$) possess a significant capability to differentiate between IDE and the standard $\Lambda$CDM model. This finding highlights the critical importance of low-redshift data in constraining the parameters of these cosmological models. Consequently, other RSD compilations that incorporate additional data within this redshift range may yield different results than those discussed above. Such variations could lead to new insights into the nature of DE and its interactions, emphasizing the need for a comprehensive analysis of low-redshift observations in future studies.

\section{Final Remarks}
\label{conclusions}

In this work, we presented updated observational constraints on a widely discussed cosmological model featuring a non-gravitational interaction between DE and DM, responsible for energy-momentum flow in the dark sector of the cosmological model. For the interaction kernel, we adopt the well-known parameterization $Q = \xi {\cal H} \rho_{\rm x}$, where the amount and direction of the energy-momentum flow are determined by the strength and sign of the coupling parameter $\xi$, respectively. These scenarios have been extensively investigated and tested using various datasets. Here, we focus on the role played by Redshift Space Distortion measurements. The novel aspects of our analysis and the main findings are summarized as follows:

\begin{itemize}

\item \textbf{New Bounds for Dark Coupling}: RSD data has a strong impact on models with $\xi < 0$, allowing us to impose very significant upper bounds in these cases. Specifically, we find that $\xi > -0.0440$ from RSD data alone. Combination with PP and DESI samples, further improves this bounds to $\xi > -0.0430$. Similar conclusions are drawn for $\xi > 0$, where we obtain the upper limit $\xi < 0.0716$ from the combination of DESI, PP, RSD, and CC data.

\item \textbf{Solution for the $S_8$ Tension}: Due to the strong correlation between $\xi$ and other key cosmological parameters such as $\Omega_m$, $H_0$, and $S_8$ (see also Figure~\ref{fig:rsd_xi_positive}), positive values of $\xi$ suppress the growth factor at low redshift ($z < 0.75$), while maintaining $\Omega_m \sim 0.31$, resulting in lower $S_8$ values compared to the $\Lambda$CDM scenario. Although IDE models featuring energy-momentum transfer from DE to DM do not resolve the Hubble tension, they do effectively mitigate the tension between CMB and RSD data concerning $\Omega_m$ and $S_8$, as shown in the right panel of Figure~\ref{fig:s8_xi}. This underscores the viability of models with $\xi > 0$ in achieving $S_8$ values that reconcile both datasets, thereby addressing the $S_8$ tension.

\item \textbf{Interaction Evidence}: Based on the statistical criteria used to select the RSD sample, we found that RSD data have a significant impact on this class of models, imposing stringent upper limits on the parameter $\xi$ across all cases analyzed in this work. Consequently, we conclude that the inclusion of RSD data strongly constrains IDE models, rendering them nearly indistinguishable from the standard $\Lambda$CDM model.
\end{itemize}

Overall, the most significant result to highlight is that IDE models with $\xi > 0$ has strong potential to resolve the tension in $S_8$, while the IDE model with $\xi < 0$ can reduce the tension in $H_0$. This dual capability suggests that a phenomenological IDE model incorporating different channels of interaction -- both positive and negative values for the coupling parameter -- could potentially resolve both tensions simultaneously, predicting high values of $H_0$ and low values of $S_8$.
In conclusion, these findings highlight that DE-DM interaction models offer a promising avenue for addressing existing tensions and provide valuable insights into the (in)consistency among different datasets. However, they also underscore the need for further theoretical and phenomenological advancements, setting the path forward to conclusively resolve these issues within a unified and consistent IDE framework that is yet to be fully understood.

\begin{acknowledgments}
\noindent The authors express their gratitude to the referee for the valuable comments and suggestions, which have contributed to enhancing the significance of the results presented in this work. M.A.S and E.S. received support from the CAPES scholarship. R.C.N. thanks the financial support from the Conselho Nacional de Desenvolvimento Científico e Tecnologico (CNPq, National Council for Scientific and Technological Development) under the project No. 304306/2022-3, and the Fundação de Amparo à Pesquisa do Estado do RS (FAPERGS, Research Support Foundation of the State of RS) for partial financial support under the project No. 23/2551-0000848-3. S.K. gratefully acknowledges the support of Startup Research Grant from Plaksha University  (File No. OOR/PU-SRG/2023-24/08), and Core Research Grant from Science and Engineering Research Board (SERB), Govt. of India (File No.~CRG/2021/004658).
E.D.V. is supported by a Royal Society Dorothy Hodgkin Research Fellowship. W.G. is supported by the Lancaster–Sheffield Consortium for Fundamental Physics under STFC grant: ST/X000621/1. 
This article is based upon work from COST Action CA21136 Addressing observational tensions in cosmology with systematics and fundamental physics (CosmoVerse) supported by COST (European Cooperation in Science and Technology).
\end{acknowledgments}

\bibliography{main}




\end{document}